\input{aipcheck}

\documentclass[
    ,final            
  ]
  {aipproc}

\layoutstyle{6x9}

\begin{document}

\title{Temporal correlator in YM$^2_3$ and \\
       reflection-positivity violation}

\author{A.R.\ Taurines}{
  address={Instituto de F\'{\i}sica Te\'orica, Universidade Estadual Paulista,
\\Rua Pamplona 145, S\~ao Paulo, SP, Brazil}
}

\author{A.\ Cucchieri}{
address={Instituto de F\'{\i}sica de S\~ao Carlos, Universidade
de S\~ao Paulo,  \\C.P. 369, 13560-970, S\~ao Carlos, SP, Brazil
} 
}

\author{T.\ Mendes}{
  address={Instituto de F\'{\i}sica de S\~ao Carlos, Universidade
de S\~ao Paulo,  \\C.P. 369, 13560-970, S\~ao Carlos, SP, Brazil
}
}

\begin{abstract}
We consider numerical data for the lattice Landau gluon propagator
obtained at very large lattice volumes in three-dimensional pure $SU(2)$
Yang-Mills gauge theory (YM$^2_3$). We find that
the temporal correlator $\,C(t)\,$ shows an oscillatory pattern and
is negative for several values of $t$. This is an explicit
violation of reflection positivity and can be related to gluon confinement.
We also obtain a good fit for this quantity in the whole time interval
using a sum of Stingl-like propagators.
\end{abstract}

\maketitle


\section{Introduction}

The reconstruction of a G\aa rding-Wightman quantum field
theory from the corresponding Euclidean Green functions is possible
only if they obey the Osterwalder-Schrader axioms. In particular,
the requirement of positive definiteness of the norm
in Hilbert space is expressed in Euclidean space
by the {\em axiom of reflection positivity}.
Introducing the {\em temporal correlator} of
a 2-point function $D(p)$ as its partial Fourier transform
\begin{equation}
C(t)\;\equiv\;D(t,0)\;=\; 
\int_{-\infty}^{\infty}\,\frac{dp_0}{2\pi}\;D(p)\;e^{ip_0\,t}\;,
\label{eq:ctdef}
\end{equation}
this axiom can be re-expressed simply as
\begin{equation}
C(t) \, > \, 0 \; .
\end{equation}
Thus, reflection positivity is violated
if this temporal correlator assumes non-positive
values.

The gluon propagator $\,D(p)\,$ is 
predicted to vanish at zero momentum
in Landau gauge \cite{Zwanziger:1991gz}.
This implies that the real-space propagator ${\widetilde D}(x-y)$
is positive and
negative in equal measure, i.e.\ reflection positivity is maximally
violated.
This result can be interpreted as one {\em manifestation of confinement}
in QCD, together with
the Kugo-Ojima confinement criterion --- which
is equivalent to the horizon condition  ---
and with the failure of the cluster decomposition (see
\cite{Alkofer:2000wg} for a detailed discussion).

An explicit violation of reflection positivity
for the Landau-gauge gluon propagator in YM$^2_3$ theory
was presented in \cite{Cucchieri:2004mf}.
Here we review briefly the results obtained there
and we try to fit the temporal correlator
$C(t)$ in the whole $t$ interval.

\section{Violation of reflection positivity}

On the lattice, the temporal correlator can be evaluated
using
\begin{equation}
C(t) =  \frac{1}{N} \sum_{k_0=0}^{N-1}
e^{- 2\,\pi\,i\, k_0\, t /N} \,D(k_0, 0) \;,
\label{eq:ltc}
\end{equation}
where $N$ is the number of points per lattice side and $D(k)$ is 
the propagator in momentum space. If the lattice action satisfies
reflection positivity, it can be shown \cite{Montvay:1994cy} that
$C(t)$ is non-negative for all values of $t$. 
Furthermore, one can consider the quantity
\begin{equation}
G(t,a) \; =\; 
\frac{1}{a^2}\left[\,C(t)\,C(t+2a) \,-\,
C(t+a)^2 \,\right] \;,
\end{equation}
where $\,a\,$ is the lattice spacing. This function is also
non-negative in the whole time interval if reflection positivity
holds \cite{Cucchieri:2004mf}.

\begin{figure}[t]
\includegraphics[height=0.65\hsize,angle=-90]{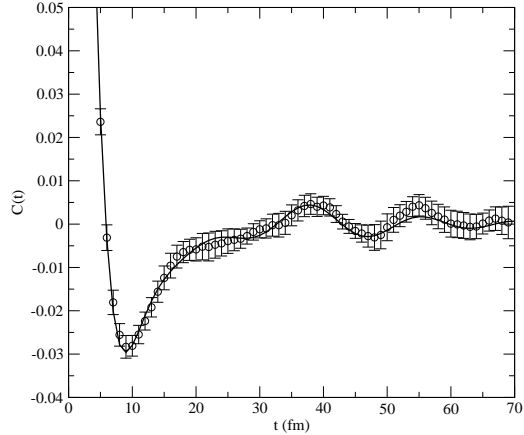}
\caption{Temporal correlator $C(t)$ as a function of $t$
for coupling $\beta=6.0$ and lattice volume $V = 140^3$.
The solid curve corresponds to the fit using the function reported in
eq.\ (\protect\ref{eq:fitStingl}).
\label{fig:redect_zoom}
}
\end{figure}

We consider the $3d$ $SU(2)$ Landau-gauge data presented in
\cite{Cucchieri:2003di},
obtained at very large lattice volumes ($V=80^3,140^3$) 
and for lattice couplings in the scaling region ($\beta=4.2,\,5.0,\,6.0$).
We find \cite{Cucchieri:2004mf}
that the temporal correlator $C(t)$ is negative
for several values of $t$, showing a clear
oscillatory behavior (see Fig.\ \ref{fig:redect_zoom}).
We also checked that $G(t,1)$ is
negative for {\em all} values of $t$ (as is the case of the
Gribov-like propagator \cite{Cucchieri:2004mf}).
Thus, we find an {\em explicit violation of positivity}
for the lattice Landau gluon propagator. Let us stress that this 
violation is clearly observable for the
three lattice couplings and for the two lattice volumes
considered.

\section{Fits for $C(t)$}

The scaling behavior of the numerical data has been
analyzed using
the matching technique described in
\cite{Cucchieri:2003di}.
We find \cite{Cucchieri:2004mf} that finite-size effects
become important only for temporal separations $t \ge 3\,fm$ 
and that all propagators become
negative at $t \approx 0.7\,fm$. In this scaling region
the temporal correlator can be well fitted
using a  sum of two Gribov-like forms \cite{Cucchieri:2004mf},
with a light-mass scale
$\,M\approx 325\,MeV$. 

Here we try to fit the data in the whole time interval using a sum
of Stingl-like propagators \cite{Aiso:1997au}
\begin{equation}
D(p^2)\;=\;\sum_{i=1}^{n} \; \frac{a_i \,p^2 \,+\,s_i}{p^4\,+\,m_i\,p^2\,+\,
                                \gamma_i^4}\; .
\label{eq:fitStingl}
\end{equation}
We find that a very good fit ($\chi/d.o.f.\sim 0.25 $) is obtained
considering $n = 3$. In Table
\ref{uniquetable} we report the fitting parameters for the
lattice volume $\,V=140^3\,$ with $\beta=6.0$.
The corresponding curve is shown in Fig.\ \ref{fig:redect_zoom}. 
Similar fits can be obtained in the other cases.

\begin{table}
\begin{tabular}{lcccc}
\hline
     &{$a_i$} & {$s_i$} &{$m_i$} & {$\gamma_i$} \\ \hline
i=1  & -0.010 &  0.0013 & -0.26  &  0.36 \\
i=2  &  0.081 &  0.0081 & -0.025 &  0.25 \\
i=3  &  0.147 & -0.013  & -0.18  &  0.36 \\ \hline
\end{tabular}
\caption{Parameters for the fitting formula (\protect\ref{eq:fitStingl}),
in lattice units.}
\label{uniquetable}
\end{table}

\begin{theacknowledgments}
This work was supported by FAPESP (Projects No.
00/05047-5 and 03/05259-0). Partial support from
CNPq is also acknowledged (AC, TM).

\end{theacknowledgments}


\bibliographystyle{aipproc}   

\bibliography{my_references}

\IfFileExists{\jobname.bbl}{}
 {\typeout{}
  \typeout{******************************************}
  \typeout{** Please run "bibtex \jobname" to optain}
  \typeout{** the bibliography and then re-run LaTeX}
  \typeout{** twice to fix the references!}
  \typeout{******************************************}
  \typeout{}
 }

\end{document}